\def\CuCN{$\kappa$-(BE\-DT\--TTF)$_2$\-Cu$_2$(CN)$_{3}$}
\def\AgCN{$\kappa$-(BE\-DT\--TTF)$_2$\-Ag$_2$(CN)$_{3}$}
\def\EtMe{$\beta^{\prime}$-EtMe$_3$\-Sb\-[Pd(dmit)$_2$]$_2$}

\def\cm{cm$^{-1}$}
\documentclass[aps,prl,twocolumn,showpacs,superscriptaddress,am]{revtex4}
\usepackage{graphicx}
\usepackage{amsmath}
\usepackage{amssymb}
\usepackage{color}
\usepackage{amsfonts}%

\begin{document} 
\title{Low-Energy Excitations in Quantum Spin Liquids Identified by Optical Spectroscopy}
\author{A. Pustogow}
\affiliation{1.~Physikalisches Institut, Universit\"{a}t
Stuttgart, Pfaffenwaldring 57, D-70550 Stuttgart Germany}
\author{Y. Saito}
\affiliation{1.~Physikalisches Institut, Universit\"{a}t
Stuttgart, Pfaffenwaldring 57, D-70550 Stuttgart Germany}
\affiliation{Department of Physics, Hokkaido University, Sapporo, Japan}
\author{E.~Zhukova}
\affiliation{Moscow Institute of Physics and Technology (State University), 141700, Dolgoprudny, Moscow Region, Russia}
\author{B. Gorshunov}
\affiliation{Moscow Institute of Physics and Technology (State University), 141700, Dolgoprudny, Moscow Region, Russia}
\author{R.\ Kato }
\affiliation{Condensed Molecular Materials Laboratory, RIKEN, Wako-shi, Saitama 351-0198, Japan}
\author{T.-H. Lee}
\affiliation{Department of Physics and National High Magnetic Field Laboratory, Florida State University, Tallahassee, Florida 32306, USA}
\author{S. Fratini}
\affiliation{Institut N\'{e}el - CNRS and Universit{\'e} Grenoble Alpes, 38042 Grenoble Cedex 9, France}
\author{V. Dobrosavljevi{\'c}}
\affiliation{Department of Physics and National High Magnetic Field Laboratory, Florida State University, Tallahassee, Florida 32306, USA}
\author{M. Dressel}
\affiliation{1.~Physikalisches Institut, Universit\"{a}t
Stuttgart, Pfaffenwaldring 57, D-70550 Stuttgart Germany}
\date{\today}
\begin{abstract}
The electrodynamic response of organic spin liquids with highly-frustrated triangular lattices has been measured in a wide energy range.
While the overall optical spectra of these Mott insulators are governed by transitions between the Hubbard bands, distinct in-gap excitations can be identified  at low temperatures and frequencies, which we attribute to the quantum-spin-liquid state.
For the strongly correlated  $\beta^{\prime}$-EtMe$_3$\-Sb\-[Pd(dmit)$_2$]$_2$,
we discover enhanced conductivity below $175~{\rm cm}^{-1}$, comparable to the energy of the magnetic coupling $J\approx 250$~K. For $\omega\rightarrow 0$
these low-frequency excitations vanish faster than the charge-carrier response subject to Mott-Hubbard correlations, resulting in a dome-shape band peaked at 100~\cm.
Possible relations to spinons, magnons and disorder are discussed.
\end{abstract}


\maketitle
%
%

Quantum spin liquids are an intriguing state of matter  \cite{Balents2010,Norman16,Savary17,Zhou17}:
although the spins interact strongly, the combination of geometrical frustration and 
quantum fluctuations prevents long-range magnetic order 
even in two and three dimensions. It took decades before clear experimental realizations of this theoretical concept \cite{Pomeranchuk41,Anderson73} became available,
first in the organic compound \CuCN, which crystallizes in a triangular pattern \cite{Shimizu03},
and later in the kagome lattice of ZnCu$_3$\-(OH)$_6$Cl$_2$
\cite{Shores05,Mendels07,Helton07}. Despite this progress, a smoking-gun experiment identifying its essential features is still lacking, and even a reliable theoretical description of real spin-liquid systems remains a subject of much dispute. 
At present, the fundamental nature of the spin-liquid state is far from being understood.

The intensely studied quantum-spin-liquid candidate Herbertsmithite (ZnCu$_3$(OH)$_6$Cl$_2$) shows no magnetic order and no indications of a spin gap down to 0.1~meV, inferring that the spin excitations form a continuum \cite{Norman16}. This important issue, however, is far from being settled
neither from the experimental nor from the theoretical side \cite{Yan11,Potter13}; the discussion on the nature of the quantum-spin-liquid state is rather controversial \cite{Kalmeyer87,Wen02,Sheng09,Qi09,Mishmash13,Motrunich05,Ran07}.

In this Letter we investigate the electrodynamic response of three organic quantum spin liquids. While close to the Mott transition the charge degrees of freedom dominate the conductivity, our optical experiments reveal considerable low-frequency absorption deep inside the Mott-insulating state. The observed dome-like feature is confined by the exchange energy $J$, suggesting a relation to the spin degrees of freedom and exotic spin-charge coupling.

We study the charge-transfer salts
\CuCN\ (abbreviated CuCN, BEDT-TTF denotes bis(ethylene\-dithio)\-tetra\-thia\-ful\-va\-lene), \AgCN\ (called AgCN) and \EtMe\ (short EtMe, here EtMe$_3$Sb stands for ethyltrimethylstibonium and dmit is 1,3-dithiole-2-thione-4,5-dithiolate), where the molecular dimers with spin-$\frac{1}{2}$ form a highly frustrated triangular lattice  \cite{Shimizu03,Itou08,Shimizu16,Pinteric16}.
At ambient pressure no indication of N{\'e}el order is observed at temperatures as low as 20~mK, despite the considerable antiferromagnetic exchange of $J\approx 220-250$~K. The origin of the spin-liquid phase is unresolved since the geometrical frustration introduced by a triangular lattice should not be sufficient to stabilize the quantum-spin-liquid state for ordinary Heisenberg exchange interactions \cite{Huse88,Capriotti99,Kaneko14}.
Recently it was proposed that, alternatively, intrinsic disorder \cite{Furukawa15,Dressel16,Lazic17} or dynamical fluctuations \cite{Yamamoto17} may play a crucial role for stabilizing the spin-liquid state in these molecular materials.

In contrast to the completely insulating material Herbertsmithite, where the on-site Coulomb repulsion $U$ and bandwidth $W$ are in the eV range ($U=8$~eV~\cite{Pustogow2017Herbertsmithite}), the energy scales of these organic compounds are significantly smaller; here, $k_B T$ has a large effect on the Mott gap already for a few hundred Kelvins as evident in dc transport~\cite{Pinteric14,Pinteric16,Lazic17}. Moreover, the molecular conductors under study are closer to the metallic phase due to weaker correlations. With $U/W \approx 1.5$, CuCN is almost at the metal-insulator transition; in fact it becomes superconducting at $T_c=3.6$~K under hydrostatic pressure of only 4~kbar \cite{Shimizu03}. For AgCN the effective correlations are more pronounced, as $U/W = 2$, while EtMe is far on the Mott insulating side with $U/W\approx 2.4$ \cite{Pustogow17}. Heat capacity measurements suggested gapless spin excitations for CuCN \cite{Yamashita08},
in contrast to thermal-transport data \cite{Yamashita09}. For EtMe both thermodynamic probes \cite{Yamashita10,Yamashita11} and magnetization data \cite{Watanabe12} favor a gapless scenario whereas NMR results indicate a nodal spin gap \cite{Itou08}.

A possible scenario featuring gapless spin excitations has been proposed by Lee and collaborators \cite{Lee05}, based on a U(1) gauge theory of the Hubbard model,  suggesting a spinon Fermi surface, which should produce metallic-like low-temperature behavior of both the specific heat and the thermal conductivity. In addition, the coupling of such spinons with an internal gauge field has been predicted to contribute to the optical conductivity \cite{Ioffe89,Ng07,Potter13} and even produce the magneto-optical Faraday effect \cite{Colbert14}. This intriguing possibility motivated our present investigations on the electrodynamic behavior of quantum spin liquids. Deep within the Mott-insulating state, where charge excitations are largely frozen out, we find signatures of a novel low-energy absorption band, which we attribute to anomalous excitations arising from the spin-disordered ground state in absence of antiferromagnetism.

From the general point of view, optical experiments are not sensitive to conventional spin excitations; however, spin-charge interactions may contribute to the low-energy optical conductivity, as suggested in theories featuring an emergent gauge field within the spin-liquid state \cite{Ng07,Zhou13}. This mechanism was predicted to produce a power-law behavior: $\sigma_1(\omega)\propto \omega^{\beta}$ with $\beta=2$ at low frequencies and a crossover to $\beta=3.3$  above $\hbar\omega_c \approx k_BT$. THz investigations on
the kagome crystal  ZnCu$_3$(OH)$_6$Cl$_2$ provided first experimental results \cite{Pilon13} analyzed in this context: at $T=4$~K a power-law behavior of the optical conductivity was obtained with an exponent $\beta = 1 - 2$ up to  1.4~THz when it crosses over into a phonon tail.
There have been several attempts to investigate also the triangular compounds with a similar scope.
Ng and Lee \cite{Ng07} analyzed the infrared spectra of CuCN, but
the underlying data  \cite{Kezsmarki06} do not extend to low-enough frequency to allow for reliable conclusions.
Subsequent infrared-reflectivity measurements by Els{\"a}sser {\it et al.} \cite{Elsasser12} could barely reach the relevant region around $\hbar \omega \approx k_BT $ down to the lowest temperatures. 

\begin{figure}
\centering
\includegraphics[width=1\columnwidth]{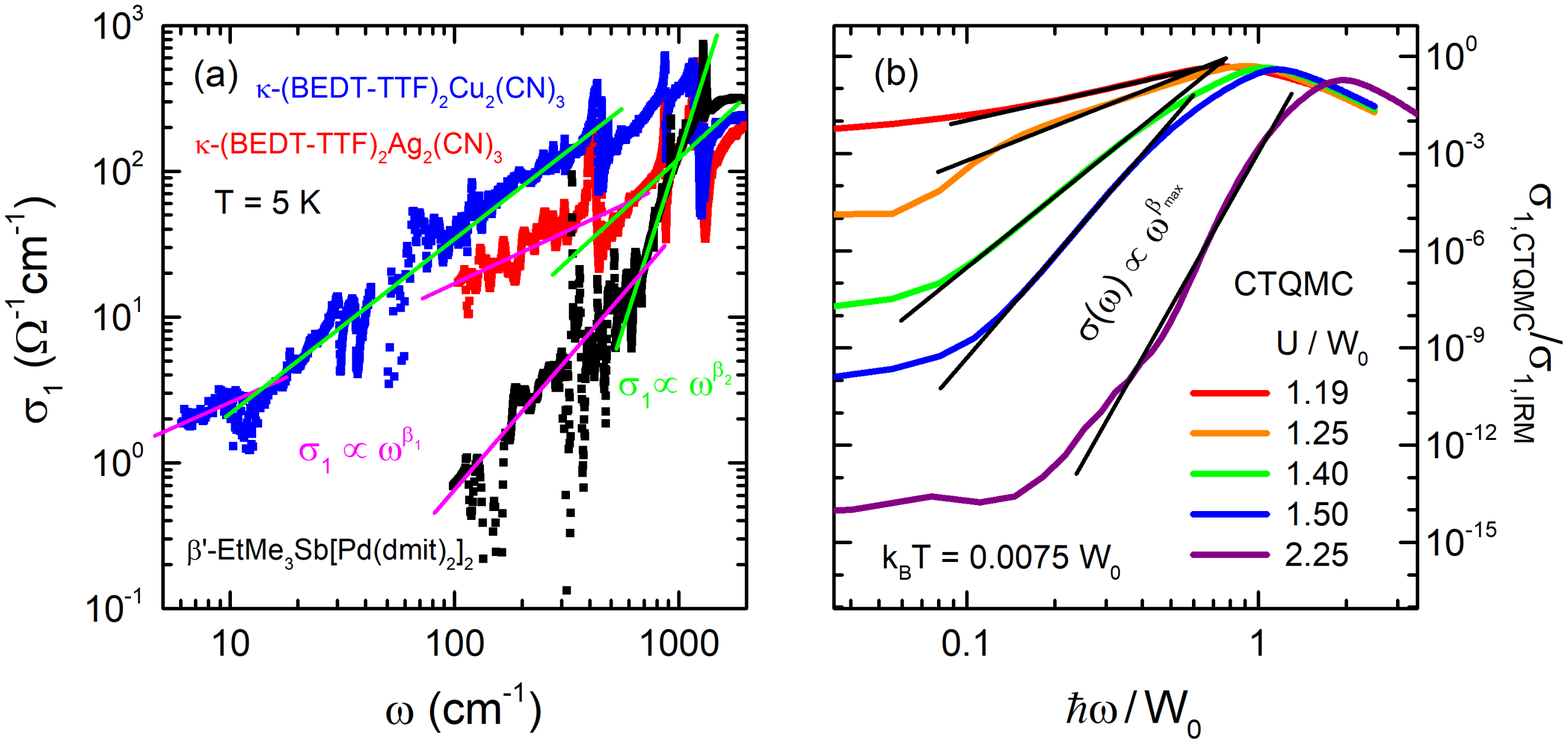}
\caption{(a)~The in-gap conductivity of \EtMe, \AgCN\ and \CuCN\ can be approximated
by power laws, as indicated by the straight lines in the double-logarithmic plot.
Two different exponents are identified in different frequency regions.
(b)~Optical conductivity calculated by dynamical mean-field theory methods employing the continuous time quantum Monte Carlo (CTQMC) quantum impurity solver for different correlation strength $U/W$ as indicated. The assumed $k_BT/W = 0.0075$ corresponds to the low-temperature limit of our experiments. $\sigma_1(\omega)$ is normalized to the Ioffe-Regel-Mott limit $\sigma_{\rm IRM}$ and levels off towards the dc conductivity at low frequencies. Consistently, the exponents increase and $\sigma_{dc}$ decreases with correlation strength.
\label{fig:PowerLaw}
}
\end{figure}
In order to provide more robust data,
we have compiled the electrodynamic properties of the dimer Mott insulators CuCN, AgCN and EtMe
with quantum-spin-liquid ground states.
The infrared optical reflectivity recorded by Fourier-transform spectroscopy
over many orders of magnitude in frequency
is supplemented by dc and dielectric studies in the kHz and MHz range and
by ellispometric measurements up to the ultraviolet in order to get more reliable extrapolations \cite{Pinteric14,Dressel16,Pinteric16,Lazic17,Pustogow17}.
Fig.~\ref{fig:PowerLaw}(a) displays the optical conductivity as obtained from the Kramers-Kronig analysis of the lowest-temperature data measured along the most conducting axes of CuCN, AgCN and EtMe.
For light polarized along the second direction of the crystal surfaces, the overall behavior is rather similar \cite{Pustogow17,Pustogow14}, indicating the two-dimensional electronic response despite a slight anisotropy.
As indicated by the green and magenta lines, we can identify a power-law behavior $\sigma_1(\omega)\propto \omega^{\beta_1}$ up to a crossover frequency $\omega_c$ above which the slope approximately doubles. The exponents depend on temperature; in the case of CuCN, for instance, they start with $\beta_{1}\approx 0.4$ and $\beta_{2}\approx 0.8$ at $T=300$~K, and then increase to 0.9 and 1.3 upon cooling, respectively.
While AgCN behaves rather similar as CuCN, the power-law exponents of EtMe are significantly larger, reaching $\beta_1 = 1.75$ and $\beta_2 = 4.2$ with a much more pronounced temperature dependence.

These observations can be explained by the increasing correlation strength $U/W$ when going from CuCN to EtMe. While metallic fluctuations dominate the in-gap absorption of CuCN,
a well-defined Mott gap forms only in EtMe upon cooling \cite{Pustogow17}.
In order to investigate the correlation dependence of the steepness of the band edge,
we calculated the optical spectra using dynamical mean-field theory (DMFT) utilizing the continuous time quantum Monte Carlo quantum impurity solver and
plot the normalized spectra in Fig.~\ref{fig:PowerLaw}(b).
As expected, the Mott-Hubbard bands
become more narrow and exhibit a steeper edge for enhanced
$U/W$, very similar to the experimental spectra in panel~(a) where the exponents are largest for the most strongly correlated compound EtMe. It should be stressed here that the DMFT approach, which focuses on local charge excitations, is not able to capture the relevant inter-site spin correlations contributing to spinon excitations. At this point we conclude that charge excitations, which are properly described by the DMFT theory, not only cause
the mid-infrared band via transitions between the Hubbard bands, but
also major parts of the sub-gap absorption with a power-law like increase in the optical range that levels off as the dc conductivity is approached.

Here we are in accord with recent theoretical results of Lee {\it et al.} \cite{Lee16}, who concluded that long-lived spinon excitations are not well-defined close to the Mott transition. As soon as the electrons become delocalized, the spin has to follow the charge movement, which destroys the coherence of the postulated spinon Fermi surface. In other words, fingerprints of spinon excitations can only be expected in the optical properties when the low-energy charge response is strongly suppressed. Among the investigated materials this is realized solely for EtMe, which is located deep inside the Mott state with a well-defined gap~\cite{Pustogow17}. Even then we have to look at small frequencies and temperatures well below the antiferromagnetic exchange coupling $J\approx 250$~K \cite{Itou08}.

\begin{figure}
\centering
\includegraphics[width=1\columnwidth]{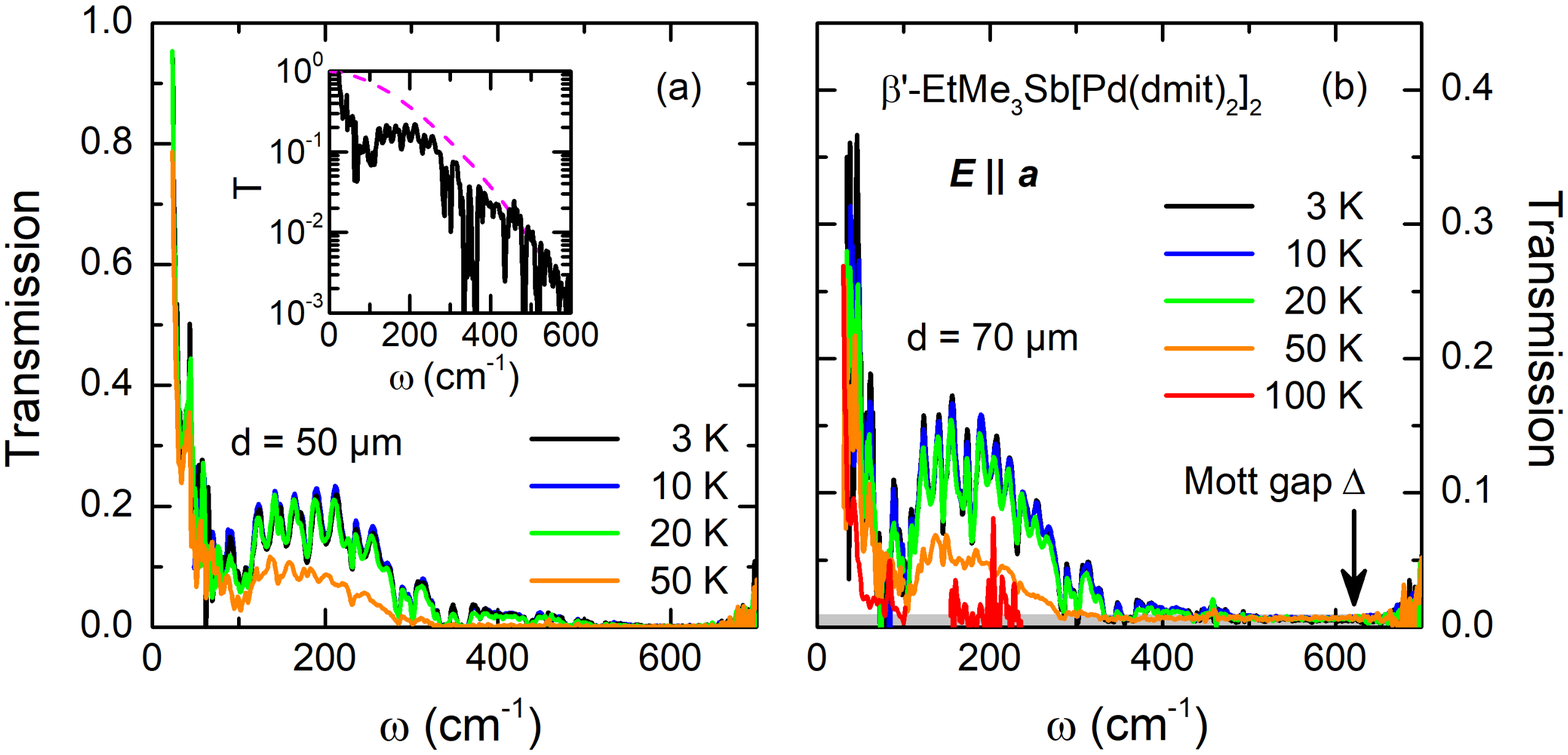}
\caption{Optical transmission spectra of \EtMe\ measured on two single
crystals of (a) 50~$\mu$m and (b) 70~$\mu$m  thickness at different temperatures.
As the Mott gap opens below $T\approx 125$~K, the transmission continuously
increases down to the lowest temperatures. The
sharp absorption features are due to molecular vibrations and lattice phonons,
best identified in the semi-logarithmic plot of the inset.
Pronounced Fabry-Perot oscillations appear at the lowest temperatures
that allow to independently determine the real and imaginary parts
of the conductivity with high accuracy. A broad absorption feature emerges
below 200~\cm. The power-law absorption $\sigma_1(\omega)\propto\omega^{1.75}$ is indicated by a dashed pink line in the inset of panel (a) where a logarithmic scale is used.
\label{fig:Trans}
}
\end{figure}
To that end, we have prepared two large EtMe single crystals as thin as 50 and 70~$\mu$m and
conducted optical transmission measurements in the low-energy range using coherent source and
time domain THz spectrometers as well as Fourier-transform interferometers.
As seen in Fig.~\ref{fig:Trans}, the specimens become successively
more transparent below the Mott gap around 600~\cm\ that opens when cooling
below 125~K.
Except of the known anisotropy, the overall behavior is similar for $E\parallel b$. 
Most important, there is a pronounced dip in the transmission centered around 100~\cm\
that indicates a novel absorption mechanism discussed in the following.

\begin{figure}[ptb]
\centering
\includegraphics[width=0.7\columnwidth]{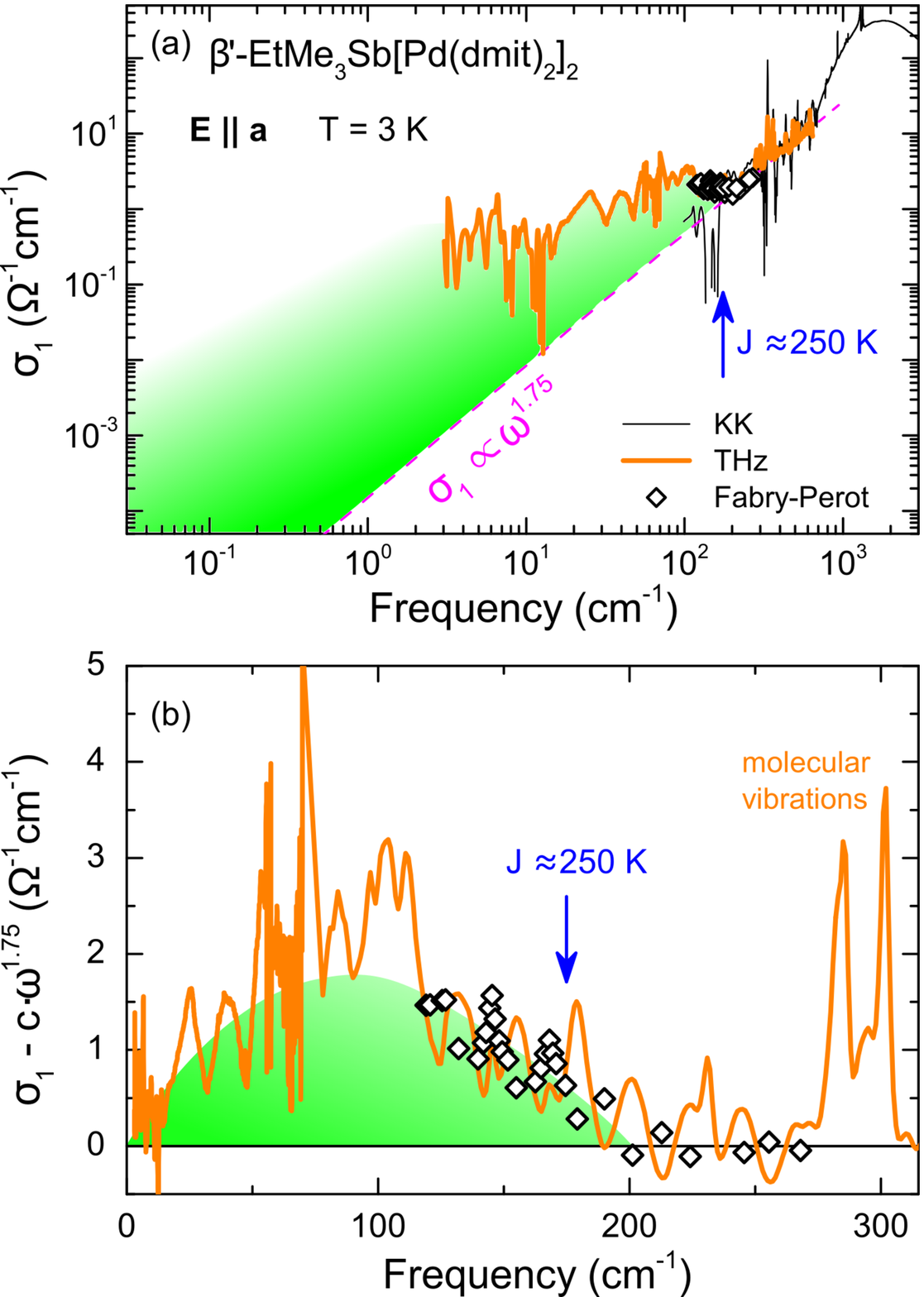}
\caption{(a)~Low-temperature conductivity spectra of \EtMe\
plotted in a broad range.
The black line corresponds to Kramers-Kronig results from optical reflectivity
measurements; it is superimposed by the data directly calculated
from transmission and reflection experiments in the THz and far-infrared ranges (orange).
The open black diamonds are obtained from fits of the Fabry-Perot oscillations.
The smooth behavior between 200 and 500~\cm\ is described by a power-law $\sigma_1(\omega) \propto \omega^{1.75}$ (dashed magenta line, cf. Figs.~\ref{fig:PowerLaw} and \ref{fig:Trans}).
As indicated in light green, the absorption exceeds the Mott-Hubbard band edge for frequencies below the antiferromagnetic exchange coupling $J\approx 250$~K.
(b)~After subtracting the power-law background related with pure charge excitations, we obtain a broad low-energy feature below approximately 200~\cm, which is close to the antiferromagnetic exchange energy of \EtMe.
Hence, we associate the green dome-like area to
low-energy excitations in the spin-liquid state.
The narrow modes at higher frequencies correspond to vibrational features.
\label{fig:Spinons}
}
\end{figure}
The real and imaginary parts of the optical conductivity are directly calculated from the optical transmission and reflection data. This way, the sensitivity is greatly enhanced and any
arbitrary extrapolation of data can be avoided.
In addition, a fit of the well-defined Fabry-Perot oscillations below 400~\cm\ allows us to independently extract the complex optical response from the transmission data only.
The so-obtained conductivity is superimposed on the results from the Kramers-Kronig analysis of the optical reflectivity and plotted in Fig.~\ref{fig:Spinons} (a). 

According to Fig.~\ref{fig:PowerLaw}, the background conductivity of EtMe
due to pure charge excitations
is given by a power law $\sigma_1(\omega)\propto \omega^{1.75}$ in the far-infrared range,
where the tail of the Mott-Hubbard band dominates.
At much lower energies it gradually levels off towards the dielectric and dc data, which are more than ten orders of magnitude smaller than the THz conductivity at liquid-helium temperatures~\cite{Lazic17,Pustogow17}, in good agreement with the DMFT results in Fig.~\ref{fig:PowerLaw} (b). Thus, the power-law behavior extrapolates down to audio and radio frequencies where the constant $\sigma_{dc}$ is approached.
In the range from 5 to 200~\cm\ the absorption measured at $T=3$~K significantly exceeds this electronic background, as indicated in Fig.~\ref{fig:Spinons} (a).
At the lowest measured frequency (3~\cm) this extra contribution is more than two orders of magnitude larger than the contribution from charge excitations and, thus, constitutes the main part of the optical conductivity. It is important to note, however, that also this novel absorption channel decays towards low frequencies, although initially not as quickly as the charge response. 

In order to unravel the nature of the exotic in-gap excitations, we subtract the $\omega^{1.75}$ power-law and thus eliminate the electronic background. Since the charge contribution is much smaller, this phenomenological procedure does not affect the low-energy part, revealing the dome-like band plotted in Fig.~\ref{fig:Spinons} (b) which peaks slightly below 100~\cm. Apart from a few small phononic features on top, the band is rather isotropic and confined to a frequency range comparable to the antiferromagnetic exchange energy  $J \approx 250~{\rm K} = 175$~\cm. Thus, we may assign this dome-shaped in-gap absorption to spin excitations, which occur when $J$ is the dominant energy scale and the electronic conductivity is sufficiently suppressed at low temperatures. 

In the measured range we do not find any indications of the power laws predicted by Lee and collaborators \cite{Lee05,Ng07}. Still, our data do not rule out $\omega^2$ behavior at lower frequency, which would be consistent with the observed decay towards zero energy: in the static limit $\sigma_1(\omega)$ decays faster
than the $\omega^{1.75}$ power-law background of the charge excitations.
In this case, spinons affect neither the optical range nor the dc response where the physics of correlated electrons prevails; they may be observed at finite temperatures in a limited frequency range well below the Mott gap. 

Another way to scrutinize the broadband spin response is via magnetic scattering techniques. While neutron scattering can map the dispersion relation of magnetic excitations~\cite{Balents2010}, Raman spectroscopy is sensitive to the spin degrees of freedom only via higher-order excitations, such as two-magnon processes~\cite{Savary17,Weber2000}. A sharp resonance is observed in antiferromagnets due to well-defined magnon energies. In quantum spin liquids, however, the absence of magnetic order gives rise to broad scattering features, commonly interpreted as the creation of a pair of spinons for each magnon~\cite{Balents2010}. Indeed, such Raman signatures were observed in EtMe, AgCN and CuCN~\cite{Nakamura2015,Nakamura2016,Nakamura2014}, as well as in the inorganic kagome and honeycomb materials Herbertsmithite~\cite{Wulferding10} and $\alpha-$RuCl$_3$~\cite{Sandilands2015}, respectively. The latter report even points out the putative fermionic behavior of the magnetic continuum at low temperatures. It is interesting to note that the Raman feature of EtMe is centered around 400~\cm\ \cite{Nakamura2015}, which is almost exactly four times larger than the peak position of the band observed here (Fig.~\ref{fig:Spinons}). Considering that a two-magnon process is equivalent to four spin-$\frac{1}{2}$ excitations, one might speculate whether spin-charge coupling directly maps spinons via the charge response. Spectroscopic studies under magnetic field may elucidate the nature of these low-energy excitations~\cite{Colbert14}. 

Although an assignment of the low-frequency band (Fig.~\ref{fig:Spinons}) to spin-$\frac{1}{2}$ excitations seems plausible, we discuss other possible sources of low-energy absorption.
Considering the rather extended width of $\approx 100$~\cm\ we can certainly rule out a simple phononic origin;
typical vibrations have a width of 5--10~\cm. A recent pressure-dependent NMR and transport study revealed the importance of intrinsic disorder for the slow dynamics of EtMe~\cite{Itou17}; still, it is unlikely that disorder alone contributes to the electrodynamic response at THz frequencies.
Having in mind the broadening effect of spin excitations via magneto-elastic coupling~\cite{Sushkov2017}, phononic processes combined with disorder may map the magnetic excitations to the electrodynamic response. Yet, the comparably strong oscillator strength of our data favors more a direct excitation process, e.g. via spin-charge interations. 

Going back to the overview on several quantum spin liquids plotted in Fig.~\ref{fig:PowerLaw}, we can now understand why
for CuCN no indications of magnetic excitations could be seen in the optical conductivity \cite{Kezsmarki06,Elsasser12,Pustogow14}. Due to the weaker correlations $U/W$ the compound is located much closer to the insulator-metal phase boundary \cite{Pustogow17} and consequently the tail of the Mott-Hubbard excitations decays much slower towards $\omega\rightarrow 0$, as depicted in Fig.~\ref{fig:PowerLaw}(b). CuCN exhibits a power-law conductivity with a weaker slope ($\beta_{1,2} = 0.4 - 0.8$) and larger absolute value compared to EtMe. Hence, the electronic contribution to the electrodynamic response of CuCN dominates well into the GHz range of frequency.


From our comprehensive studies of three organic compounds with a triangular lattice
we learnt that the large electronic conductivity even below the Mott-Hubbard gap
makes it difficult to identify electronic excitations of the quantum-spin-liquid state.
Only when investigating the strongly correlated Mott insulator \EtMe\ at
very low frequencies and temperatures, we succeeded identifying an excess conductivity
that cannot be explained by the charge response of the correlated electrons.
Upon subtracting their smooth power-law background, a broad mode is identified
delimited by $J$ at its high-energy end; the strong decrease for $\omega\rightarrow 0$
is consistent with the $\omega^2$ dependence expected for spinons.
 This result is in excellent
agreement with the recent picture based on a spinon-extension of dynamical mean field theory \cite{Lee16}, which argue that the controversially discussed spinon Fermi surface melts away upon approaching the Mott metal-insulator transition.
As soon as the electrons start moving as a whole, spin-charge separation is lost and the spin excitations are not independent from the charge motion any more. We suggest that it is worth turning back to Herbertsmithite and its recently synthesized analogues~\cite{Sun2016,Puphal2017}, which exhibit a more pronounced charge gap~\cite{Pustogow2017Herbertsmithite}, and investigate these strongly correlated compounds on a kagome lattice at extremely low temperatures and frequencies.

\acknowledgments
We thank P.A. Lee, S. Tomi{\'c} and R. Valent\'i for valuable discussions. The project was supported by the Deutsche
Forschungsgemeinschaft (DFG, Grants No. DR228/39-1
and No. DR228/51-1), the Russian Ministry of Education
and Science (Program 5 top 100) and the MIPT grant for
visiting professors. R. K. was supported in part by the
Japan Society for the Promotion of Science Grants-in-Aid
for Scientific Research (Grant No. 16H06346). Work in
Florida was supported by the NSF Grant No. DMR-
1410132, and the National High Magnetic Field
Laboratory through the NSF Cooperative Agreement
No. DMR-1157490 and the State of Florida.

\end{document}